\documentclass{aa}
\usepackage{psfig}
\usepackage[comma,authoryear]{natbib}
\usepackage[latin1]{inputenc} 

\begin{document}

\title{Tracking granules at the Sun's surface and reconstructing velocity
fields. II. Error analysis}

\titlerunning{Source of errors in the CST algorithm}

\author{R. Tkaczuk\inst{1}, M. Rieutord\inst{1}, N. Meunier\inst{2}, T.
Roudier\inst{2}}

\authorrunning{Tkaczuk et al.}

\institute{Laboratoire d'Astrophysique de Toulouse et Tarbes,
UMR 5572, CNRS et Université Paul Sabatier Toulouse 3, 14 avenue E. Belin, 31400 Toulouse,
France
\and Laboratoire d'Astrophysique de Toulouse et Tarbes, UMR
5572, CNRS et Université Paul Sabatier Toulouse 3,
57 Avenue d'Azereix, BP
826, 65008 Tarbes Cedex, France\\
\email{tkaczuk@ast.obs-mip.fr, rieutord@ast.obs-mip.fr,
meunier@ast.obs-mip.fr, roudier@ast.obs-mip.fr}
}

\offprints{M. Rieutord}

\date{Received ; Accepted}

\abstract{
The determination of horizontal velocity fields at the solar surface is
crucial to understanding the dynamics and magnetism of the convection zone
of the sun. These measurements can be done by tracking granules.}
{
Tracking granules from ground-based observations, however, suffers from
the Earth's atmospheric turbulence, which induces image distortion. The focus
of this paper is to evaluate the influence of this noise on the
maps of velocity fields.}
{
We use the coherent structure tracking algorithm developed recently and
apply it to two independent series of images that contain the same
solar signal.}
{
We first show that a $k-\omega$ filtering of the times series of images
is highly recommended as a pre-processing to decrease the noise, while,
in contrast, using destretching should be avoided. We also
demonstrate that the lifetime of granules has a strong influence on the error
bars of velocities and that a threshold on the lifetime should be imposed to
minimize errors. Finally, although solar flow patterns are easily
recognizable and image quality is very good, it turns out that a time
sampling of two images every 21~s is not frequent enough, since image
distortion still pollutes velocity fields at a 30\% level on the
2500~km scale, i.e. the scale on which granules start to behave like
passive scalars.}
{
The coherent structure tracking algorithm is a useful tool for noise
control on the measurement of surface horizontal solar velocity fields
when at least two independent series are available.}


\maketitle

\section{Introduction}

Movies of the solar surface show that it is a place of intense turbulent
fluid flows where three major scales (granulation, mesogranulation, and
supergranulation) have been pointed out.  In order to better understand
the underlying dynamics, it is crucial to be able to measure the velocity
fields. As far as horizontal flows are concerned, the basic techniques have
relied on measuring the displacement of granules that, as shown
by \cite{RRLNS01}, trace the fluid flows on scales larger than 2.5~Mm.

Two algorithms have been devised to transform a time sequence of images
into a sequence of horizontal velocity fields maps. These are the LCT
algorithm (i.e. local correlation tracking, see \citealt{NS88}) and the
CST algorithm (coherent structure tracking, see \citealt{RRRD07}
hereafter referred to as Paper I).

When the surface velocity field is known, one is usually interested in
identifying/following the dynamical structures of the flow - like
vortices, upwellings, or downwellings. The identification of these
structures demands, however, computing the velocity gradients
like the divergence or the vorticity. In Paper I, it has been pointed out
that such quantities are very sensitive to the noise induced
by terrestrial atmospheric distortion, since they are derivatives of the
velocity field. The use of the velocity field to describe the dynamics of
the solar surface thus needs to be complemented by an error analysis that
both evaluates the significance of the observed dynamical features and
gives a way to eliminate, or at least reduce, the impact of errors.

The aim of this paper is to analyse the consequences of errors in the
final result of velocity, vorticity, and divergence fields. As mentioned
above, the main source of errors comes from the distortion of images
induced by the Earth's atmospheric turbulence. The case of errors or,
equivalently, the precision of measurements has already been discussed
for the LCT algorithm by \cite{NS88}, who mentioned errors the order
of 20~m/s on the velocity field. Further work by \cite{SBNSS95}
showed that this precision was certainly largely overestimated. More
recently, \cite{PBD03} investigated the case of interpolation errors,
associated with the LCT algorithm, which also spoil the final
result. Here, we focus on the CST algorithm and try to give a neat
picture of the influence of the Earth's atmospheric distortion on the
measurement of the velocity fields on different scales.

We organised the paper as follows. Using two independent series of
images of the solar surface, we first evaluate the noise induced on the
positions of the granules and how image pre-processing can reduce it.
We then focus on the way the noise influences the final velocity fields
on different scales and analyse its propagation up to the curl and
divergence maps.  Conclusions and outlooks follow.

\begin{figure} 
\centerline{
\psfig{figure=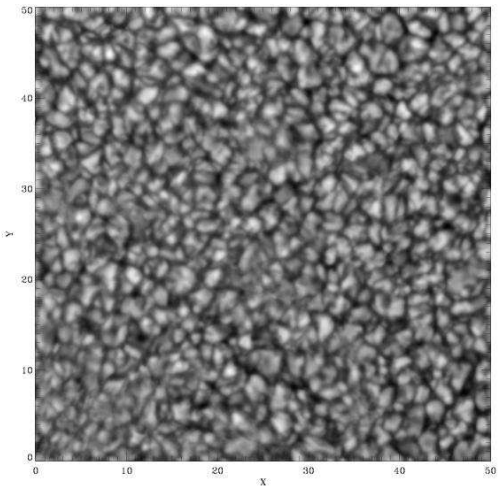,width=7.0cm}}
\caption{View of the region used for the tests. X and Y scales are in
arcsec.}
\label{exemple}
\end{figure}

\begin{figure} 
\centerline{\psfig{figure=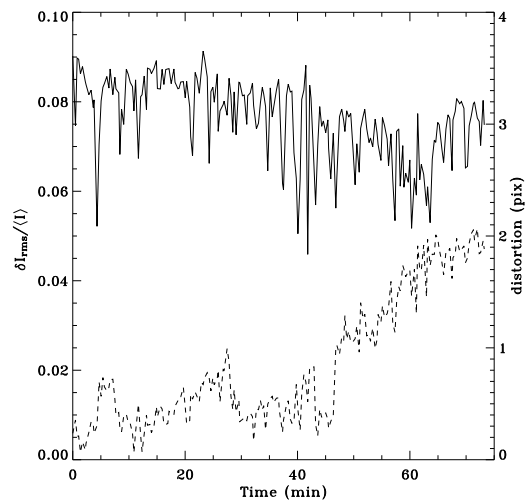,width=7.0cm}}
\caption{Indices of image quality. Contrast (solid line)
and distortion amplitude (dashed line, scale on right). Distortion is
the rms displacement, evaluated with local correlation tracking, between
two images of a pair.}
\label{image_quality}
\end{figure}

\begin{figure} 
\centerline{\psfig{figure=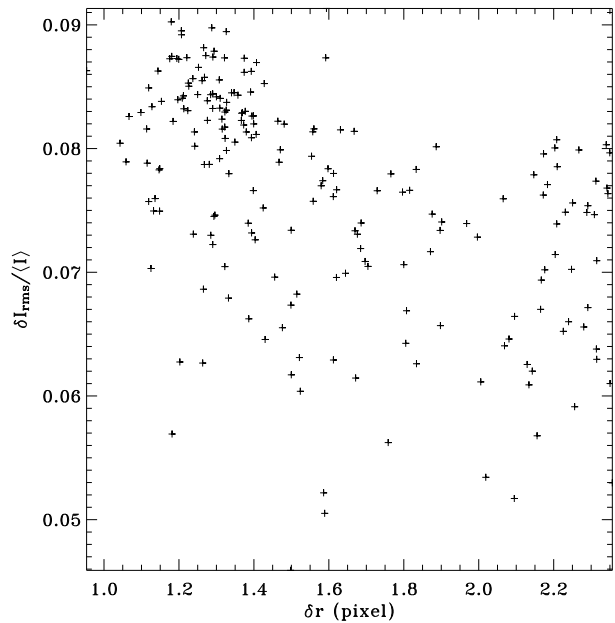,width=8.8cm} }
\caption{The normalized intensity contrast versus the rms displacement
of granules between the two images of each pair.}
\label{erreur_vs_rms}
\end{figure}

\begin{figure} 
\centerline{\psfig{figure=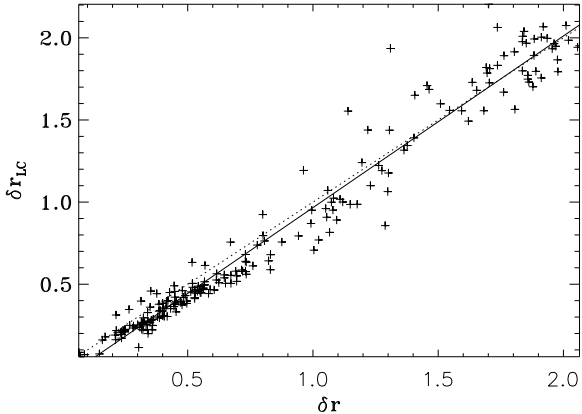,width=8.8cm}}
\caption{Mean distance between granule centres of gravity for each image pair 
versus the mean displacement due to atmospheric turbulence estimated by 
local correlation (one point per image pair). Both displacements are in
pixels. The solid line shows the line of 
equation $y=x$ and the dotted line is a linear fit over the points.
}
\label{erreur_vs_turb}
\end{figure}

\section{Observations}

We use a time series of images obtained on June 5, 1993 at
the SVST (Swedish Vacuum Solar Telescope), Observatorio Roque
de los Muchachos, La Palma (data courtesy of P. Brandt, see also
\citealt{SBN94}). This data set has already been studied by many authors
(\citealt{SBS97,SBS97b}; \citealt{SBS99,SVBHH99}; \citealt{DSBS02}; \citealt{GB02};
\citealt{RLRBM03}). The original
series consists of 1868 image pairs of size 1310$\times$970 pixels taken
at $\lambda=468\pm5$~nm.  The time between two pairs of images is close
to 21 seconds. Images from a pair are separated by a few seconds (3~s on
average and always less than 14~s). The pixel size is 0\farcs125, and
the spatial resolution is near the diffraction limit 0\farcs25.  The field
of view is  2\farcm7$ \times $2\farcm0. However, the instrument leads
to a rotation of the field of view, and the area observed on the Sun
at the beginning of the time series is different from the one at the
end. The rotation centre is located on a pore at pixel coordinates $\left(
590,102 \right)$ (not in the field-of-view used here).

For our investigations described below, we used a subsample of 210
pairs of frames (images 841 to 1262), covering $\sim$~77 minutes. In
Sect.~6, slightly more data have been used (images 800 to 1298
corresponding to a $\sim$~87~min sequence).
As shown in Fig.~\ref{exemple}, we extract a region of $401 \times 401$
pixels centred on a (magnetically) quiet zone.

The main advantage of this data set is that it contains two independent
sequences of images that can be considered as representing the same
solar signal. The only difference comes from the Earth's atmospheric
distortion, the effects of which can thus be analysed.  Moreover,
as shown by Fig.~\ref{image_quality}, image quality varies during the
sequence. Although the contrast remains almost constant, we see that the
amplitude of distortion increases after t=45~min. We thus have at our
disposal a ``good" sequence where mean distortion is about 0.5 pixel and a
``bad" sequence, where mean distortion reaches an amplitude of 2 pixels.
In the following when we refer to ``good" data, we mean the first
45~min, while ``bad" data will designate the remaining sequence.

\section{Sources of errors in CST}

In the CST method, errors are introduced through the segmentation
(i.e. through determination of the granule positions)
and through interpolation done in order to obtain a velocity field
sampled over a regularly-spaced grid.  The precision at which the velocity
field can be measured thus depends on:

\begin{itemize}
\item the precision at which a granule position is determined,
\item the duration of granule tracking,
\item the temporal resolution (time interval between velocity maps),
\item the spatial resolution of the grid used to sample the velocity
field.
\end{itemize}
This list shows that the CST algorithm allows us to identify all the
crucial steps and thus to follow the propagation of errors from the
beginning to the end of the computation.

\section{Standard deviation of granule positions}

The first step in the error analysis is to estimate the error on the granule 
positions\footnote{By granule position we mean the position of the
centre of gravity of the granule in the segmented image.}. We first apply
it to the raw data, and then use it to estimate the performances of
pre-processing steps that can be performed before granule segmentation,
namely on the image sequences.

\subsection{Method}

To estimate the standard deviation on granule
positions, we use pairs of images. Because of their quasi-simultaneity,
they represent the same solar surface, and
all the differences between them come from atmospheric 
distortions, instrumental effects, or processes applied on the data. 

After the segmentation step, the algorithm identifies granules present
in both images of the pair. We then simply measure the distance between
two identical granules. More precisely, we consider that the position of
a granule is controlled by two random variables $(x_i,y_i)$ for which we
have two realisations. We thus construct two other random variables
$(\delta x_i,\delta y_i)=(x_{i,1}-x_{i,2},y_{i,1}-y_{i,2})$, whose
variance is just twice that of $(x_i,y_i)$. Thus with a pair of images
we can estimate the mean error over the field of view in each direction
$x$ and $y$ by

\[ \sigma_x=\sqrt{\frac{1}{2N}\sum_{i=1}^N (x_{i,1}-x_{i,2})^2}, \;
\sigma_y=\sqrt{\frac{1}{2N}\sum_{i=1}^N (y_{i,1}-y_{i,2})^2},\]
and the mean displacement
$\delta r = \sqrt{\sigma_x^2+\sigma_y^2}$.

These quantities are clearly estimates of the image quality through
atmospheric distortion. Interestingly enough, we compared this estimate
with the intensity contrast usually used to indicate image quality. As
shown in Fig.~\ref{erreur_vs_rms}, the correlation between the two indicators is
rather poor, meaning that they are largely decoupled although they both
come from atmospheric turbulence! We interpret this result,
tentatively, as showing that different layers of the Earth's
atmosphere control the contrast and the distortion.

Furthermore, we compared (see Fig.~\ref{erreur_vs_turb}) the displacement
of granules with the displacement field derived from a local correlation
tracking (using an FWHM of 20 pixels) between the two images of a pair. As
expected, the correlation is much better ($\sim0.96$).

\subsection{Application to pre-processing}

One of the interesting applications of the error estimate on the
granule position is to allow the evaluation of the effectiveness of
various pre-processings applied to the images before estimating the
granule position. A measure of $\sigma_x$ before and after a given
pre-processing indicates its performance. We have applied this
approach to the $k-\omega$ filtering and to the destretching.

\begin{figure}[ht]
\hspace{0cm}\psfig{figure=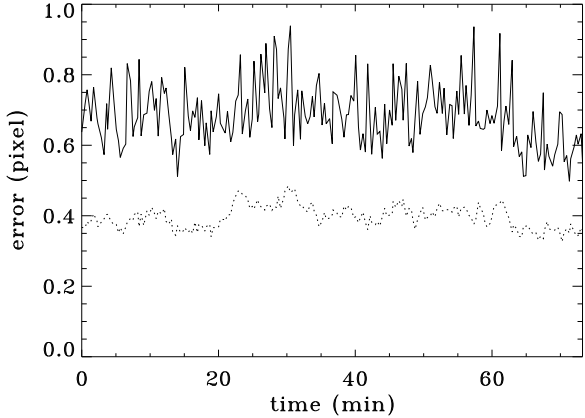,width=8.8cm}
\caption{Estimation of the error $\sigma_x$ on the determination of the centre
of gravity position in the $x$ direction as a function of time
with no $k-\omega$ filtering (solid line) and with the filtering
(dotted line).}
\label{err_kw}
\end{figure}

\begin{table}
\begin{center}
\begin{tabular}{cccc}
\hline \hline
\multicolumn{4}{c}{Good data} \\
& $\sigma_x$& $\sigma_y$ & $N_g$ \\
No processing &$0.696 \pm 0.081$&$0.701 \pm 0.084$&$ 1077 \pm 91$\\
$k-\omega$ & $0.416\pm0.037$&$0.418 \pm0.028$&$ 1096 \pm 34 $\\
\hline
\multicolumn{4}{c}{Bad data} \\
& $\sigma_x$& $\sigma_y$ & $N_g$ \\
No processing &$0.694 \pm 0.075$&$0.727 \pm 0.079$&$ 803 \pm 150$\\
$k-\omega$ & $0.470\pm0.039$&$0.527 \pm0.045$&$ 972 \pm 89 $\\
\hline \hline
\end{tabular}
\caption{Comparison of the performances obtained on time series with
high and low quality images with and without $k-\omega$ filtering.
}
\label{tabperfgood}
\end{center}
\end{table}

\begin{figure}[ht]
\psfig{figure=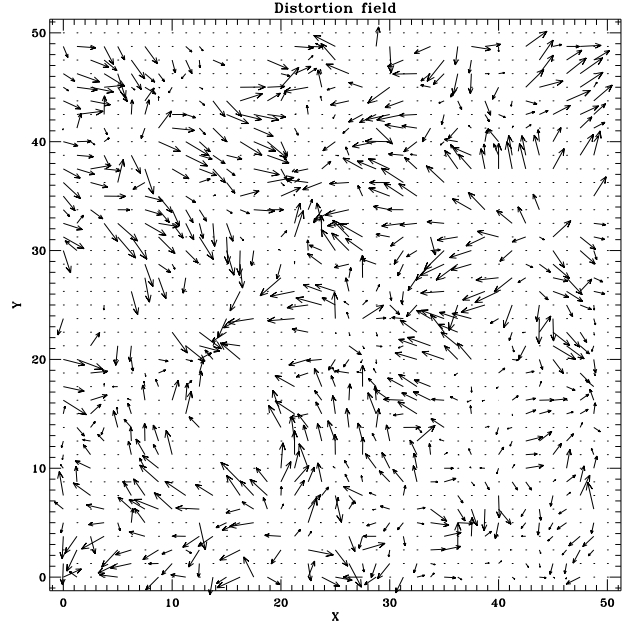,width=8.8cm,height=8.45cm}
\caption{An example of a map of displacements due to the Earth's atmospheric
distortion as sampled by granules. Regions without data have too low
an image quality for a granule to be identified. X and Y are in arcsec.}
\label{distor_map}
\end{figure}

\begin{figure} 
\hspace{0cm}\psfig{figure=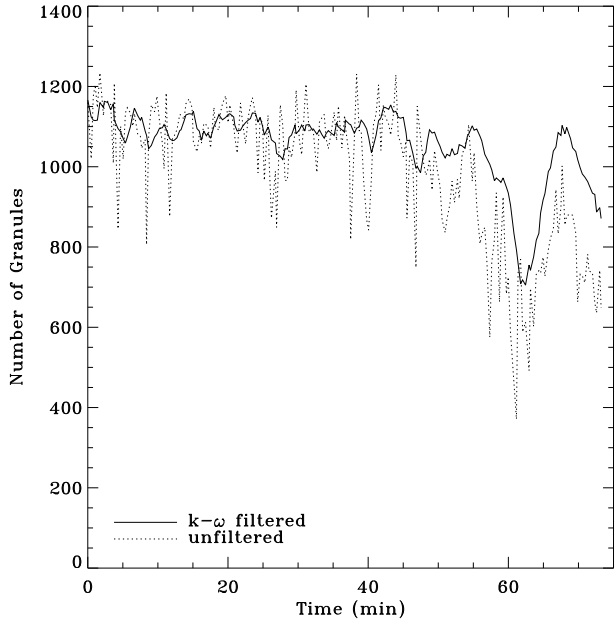,width=8.8cm}
\caption{Comparison of the number of granules for each image
for raw data (dotted line) and for $k-\omega$ filtered data (solid
line). The number of granules is defined as the number common
to both images taken 'simultaneously'. We note that filtering attenuates
the fluctuations and when atmospheric distortion increases, filtering
reduces the losses of granules.}
\label{nbg_rc_vs_dtpkw}
\end{figure}

\begin{figure}[ht]
\psfig{figure=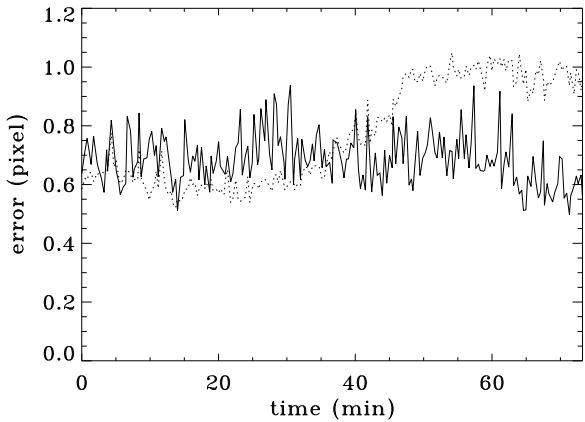,width=8.8cm}
\caption{Same as Fig.~\ref{err_kw} but for destretching; the solid line is
for unprocessed data, while dotted lines show the error ($\sigma_x$) with
the destretched sequence.}
\label{err_destr}
\end{figure}

\begin{figure}[ht]
\psfig{figure=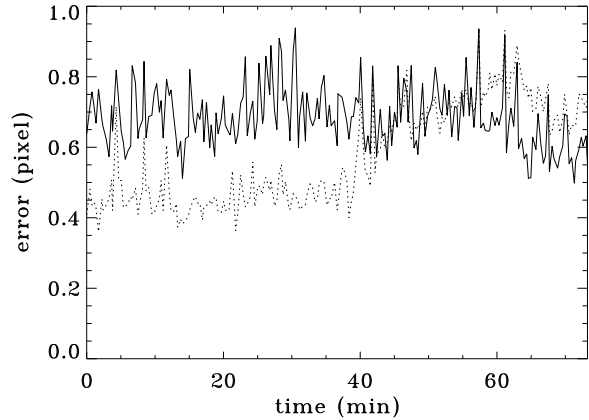,width=8.8cm}
\caption{Efficiency of destretching: the dotted line shows the
difference between two simultaneous images when one has been destretched
with the other as reference.}
\label{zero-err}
\end{figure}

\subsubsection{$k-\omega$ filtering}

The $k-\omega$ filtering \cite[see for example][]{TTT87} 
acts both spatially and temporally. In practice, it is a thresholding in 
the Fourier space. The contributions of all frequencies corresponding to 
a phase velocity higher than the threshold are eliminated.  

Let us note that the data we use are irregularly sampled.  The time step
between two pairs of images is about 21 seconds, within a few seconds
(the maximum deviation with respect to a periodic sampling is 15~s,
while the average deviation is 6.3~s). The
same data set has been used by \cite{DSBS02}, who interpolated them to get
a regularly spaced time series.  We consider that this is an unnecessary
refinement since, within a few seconds, the solar signal does not
change. Hence, with a regular time step, images would differ from ours
by just a different realisation of the Earth's atmospheric turbulence.

Figure~\ref{err_kw} shows the average pixel error on the $x$-position
of the granules for data pre-processed with a $k-\omega$ filtering and
data without pre-processing. We used a threshold for the phase velocity
of 4~km/s. The figure shows that the performance is quite 
improved over the whole time series by the use of this filtering. The error 
for the $y$-component is within an order of magnitude of the error 
for the $x$-component.

One may wonder why the increasing distortion seen after t=45~min in
Fig.~\ref{image_quality} does not appear in Fig.~\ref{err_kw}. This comes
from the way the error is measured. Indeed, only granules close enough in a
given pair of images are kept. The distortion map in Fig.~\ref{distor_map}
shows the lack of measurements in some regions: image quality is not good
enough for granules to be identified from one frame to the other. With
this method it is clear that high-amplitude distortion does not show up
with increased error in granule position, but instead appears in a reduction
of the number of ``valid granules". In Fig.~\ref{nbg_rc_vs_dtpkw} we
see the loss of granules when conditions deteriorate and how $k-\omega$
filtering improves the situation. Numbers in Table~\ref{tabperfgood}
also illustrate this process.  The errors only slightly increase from the
``good" to the ``bad" data, but the number of granules, and in other words
the number of measuring points, is reduced by 20\% without filtering,
while the loss is 10\% with filtering.

\subsubsection{Destretching}

This method, introduced by \cite{NS88}, is based on the same principles
as the LCT. It uses a local correlation scheme to determine the displacements
with respect to a reference image. From these displacements, images
are stretched by interpolation in order to be the closest possible
to the reference image. This method aims at correcting the effects of
atmospheric turbulence and allows us to  compare images that are then
plagued with the same distortion.

The pre-processing we used applies 4 successive corrections based on
local correlation tracking with the following parameters: an FWHM of 31 pixels
with a 62 pixel step, an FWHM of 62 pixels and a 31-pixel step, a 32
FWHM and a 16-pixel step, and then an FWHM of 20 pixels and a 10-pixel
step. This combination has been established empirically and seems to
give the best results.  We remind that 62 pixels correspond to
7.75\arcsec\ and 10 pixels to 1.25\arcsec.

Figure~\ref{err_destr} compares the error estimate on the granule positions in 
the $x$-direction between the destretched and raw time series.
Note that when image quality is good ($t\leq45$~min) errors remain of
the same order of magnitude as in raw data but fluctuations are less
important, while when image quality decreases ($t>45$~min), errors
increase. Quite clearly the destretched series does not compare
favourably with the raw one. On the contrary, destretching seems to
worsen the results when the image quality is slightly degraded.

Actually, the poor performance of destretching on error reduction could
be anticipated from the result displayed in Fig.~\ref{zero-err}. There
we plot the mean distance between the granules in two simultaneous
plates when one plate has been destretched to the other. If destretching
were perfect, the error would vanish. Clearly, this is not the case:
in the sequence with 'intense' atmospheric turbulence the error is
not reduced at all, while during the good sequence, a small factor
1.5 is gained. Therefore we interpret the error increase generated by
destretching as evidence that the destretching process decorrelates
from the true displacement of granules and thus introduces a new random
variable whose dispersion adds to the original signal as shown by the
factor $\sqrt{2}$ taken by the error. This decorrelation may probably
come from a change in scale of the distortion motions that is no longer
matched by the destretching process optimized for the first frames
of the sequence. If this interpretation is correct, the destretching
process would need a readjustment of the local correlation tracking
parameters from time to time, making the whole processing very costly
computationally. The only good point introduced by destretching is the
reduction of the fluctuations.

\subsubsection{Distortion noise on different scales}

We have shown in Paper I that multi-resolution analysis was an
interesting tool for the determination of flow structures. We may thus
wonder how the distortion noise affects the different scales and
therefore resists to wavelet filtering. Using maps such as the one in
Fig.~\ref{distor_map} we compute the different components of a
multi-resolution decomposition as will be used in the analysis of the
velocity field. Quantitatively we show in Table~\ref{distorscale} the
values of the distortion amplitude on different scales. This table again shows
the importance of $k-\omega$ filtering in the reduction of errors,
especially on large scales.

\begin{table}
\begin{center}
\begin{tabular}{ccc}
\hline \hline
Resolution & $\sigma_x$ & $\sigma_x(k-\omega)$ \\
\hline
1\farcs25 & 0.72 & 0.42 \\
2\farcs50 & 0.37 & 0.18 \\ 
5\farcs00 & 0.28 & 0.11 \\
10\farcs0 & 0.20 & 0.08 \\
\hline \hline
\end{tabular}
\caption{Rms fluctuations of the position of granules on a typical pair
of images shown for different resolutions with the multi-resolution
analysis and with and without $k-\omega$ filtering. $\sigma$s are given
in image pixel. The 1\farcs25 resolution corresponds to unfiltered data.
}
\label{distorscale}
\end{center}
\end{table}

\subsubsection{Conclusion}

To conclude this section, two points should be underlined. On the one
hand, destretching is certainly an unnecessary complication whose only
positive effect is to reduce the noise fluctuations when correctly tuned
to the distortion scales; otherwise, it is likely to double the variance
of the signal. On the other hand, $k-\omega$ filtering appears as the
required pre-processing whose effect on noise reduction is clear. We
surmise that, when using time-sequences with a higher time-sampling, the
noise reduction by $k-\omega$ filtering will be even more efficient.

\begin{figure} 
\hspace{0cm}\psfig{figure=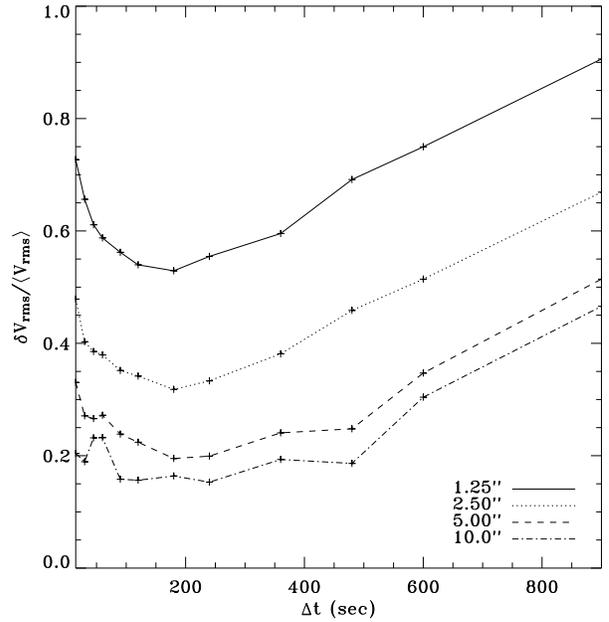,width=8.8cm}
\caption{Error on the x-component of the velocity as a function of the
threshold on the granule lifetime. The different lines are for different
resolutions of the multi-resolution decomposition; the solid line shows
the raw case (no wavelet filtering). The optimal threshold clearly
appears around 3~min.}
\label{err_fct_seuil}
\end{figure}

\begin{figure}
\hspace{0cm}\psfig{figure=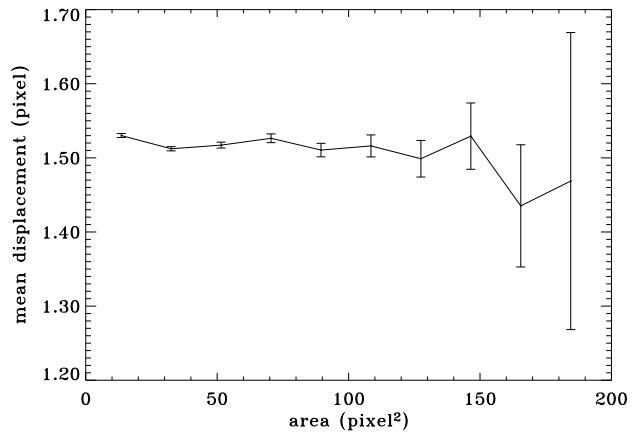,width=8.8cm}
\caption{Average distance between the granule centre of gravity
for a pair of images as a function of the granule size.}
\label{err_vs_surf}
\end{figure}

\section{Propagation of errors from granule positions to the velocities}

After examining the estimation of errors on the granule position in the
previous section, we now study the propagation of these errors to the velocity
of the granules, after averaging in time and over a given spatial range.

\subsection{Error propagation}

As seen above, the first step of the velocity computation in the CST
algorithm is to determine of the position of granules on each
image.
The duration of the tracking of the $k^{\rm th}$ granule is given
by $\Delta t_k = t_{n_f(k)}-t_{n_i(k)}$, where $n_i(k)$ 
is the image where the granule $k$ appears and $n_f(k)$ the image
where it disappears.

Over an image, granules are much less dense than pixels, making the
velocity field sampled on a much coarser grid whose elements
have size $\delta$. The velocity at a grid coordinate $(x,y)$ is
assumed to be the average of the velocity of granules whose average
coordinates  belong to the domain $D$  around $(x,y)$ (i.e. in
$[x-\delta/2,x+\delta/2],[y-\delta/2,y+\delta/2]$ ). The spatial
resolution of the velocity field is given by the mesh size $\delta$.
The $x$-component of the average velocity in $D$ is given by

$$ V_{x_p}=\frac{1}{N}\sum_{k \in D} V_{x_k} = \frac{1}{N}\sum_{k \in D}
\frac{x_{k,n_f(k)}-x_{k,n_i(k)}}{\Delta t_k}  $$
where $N$ is the number of trajectories falling in $D$. The uncertainty
on the value of $V_{x_p}$ is then

$$ \sigma_{V_{x_p}}^2= \frac{1}{N^2}\sum_{k \in D}
\frac{\sigma_{k,n_i}^2+\sigma_{k,n_f}^2}{\Delta t_k^2}. $$

If we assume that the dispersion on the centre of gravity remains the
same for the whole time series ($\sigma_x$) and that the time interval
$\Delta t_k$ is the same for all granules (i.e. all granules have the
same lifetime $\Delta t$), then the expression of the dispersion of the
average velocity in  $x$ is

$$ \sigma_{V_{x_p}}^2=\frac{1}{N^2}\sum_{i \in
D}\frac{2\sigma_x^2}{\Delta t^2} = \frac{2\sigma_x^2}{N\Delta t^2}. $$
In this case, the error on the velocity varies like

\begin{equation}
\delta V=\frac{\sqrt{2}\sigma_x}{\sqrt{N}\Delta t}.\label{preci}
\end{equation}

We see that precise velocity values need many granules in a grid element and a
long time interval. In other words, errors are less if a coarse
resolution in space and time is used. A trade off must be found (see
below), but it is clear that this technique will be more appropriate to slowly
evolving large-scale flows than to rapidly varying small-scale ones.

\begin{figure}[ht]
\centerline{\psfig{figure=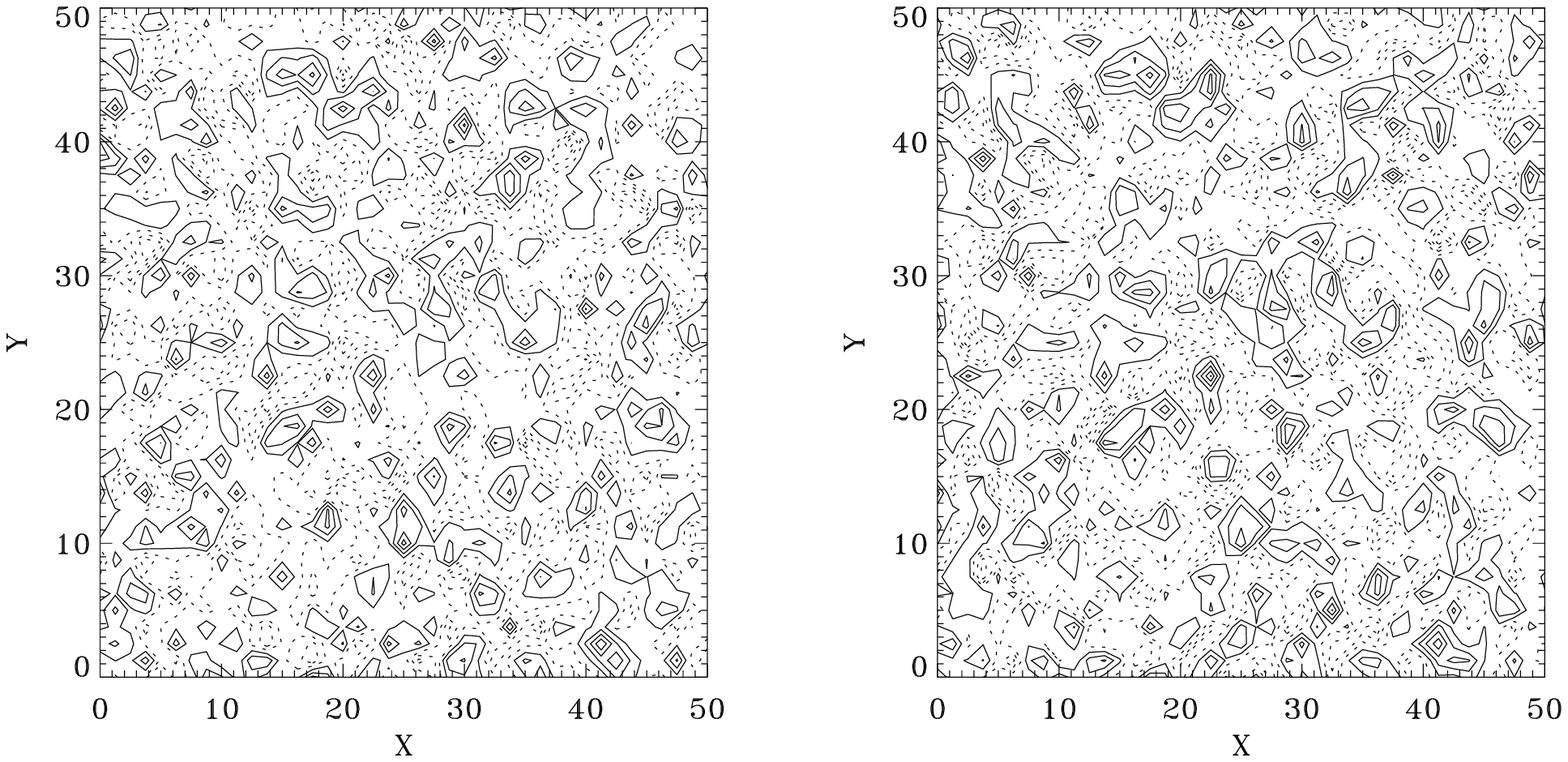,width=9.5cm,height=4.4cm}}
\centerline{\psfig{figure=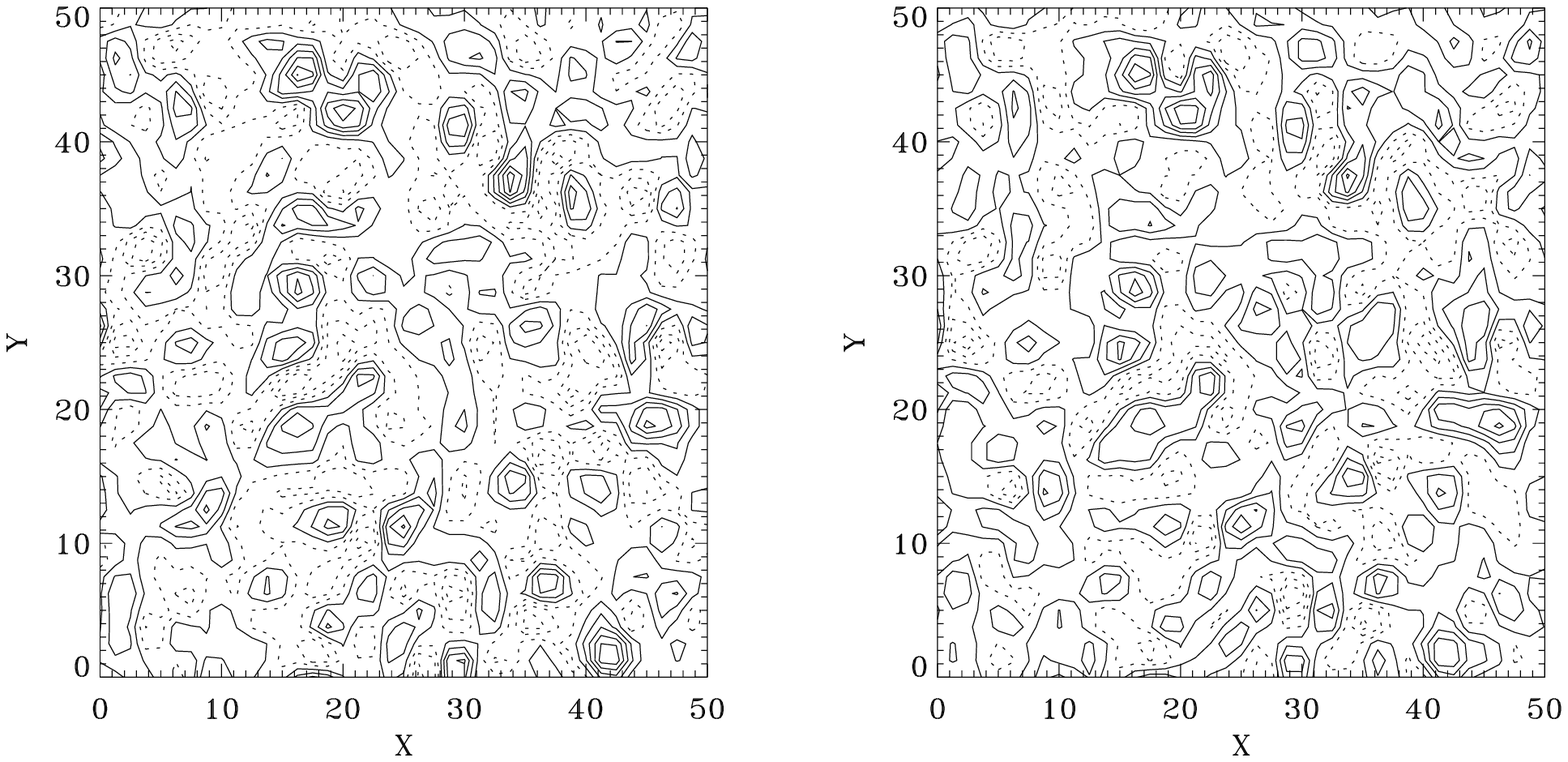,width=9.5cm,height=4.4cm}}
\centerline{\psfig{figure=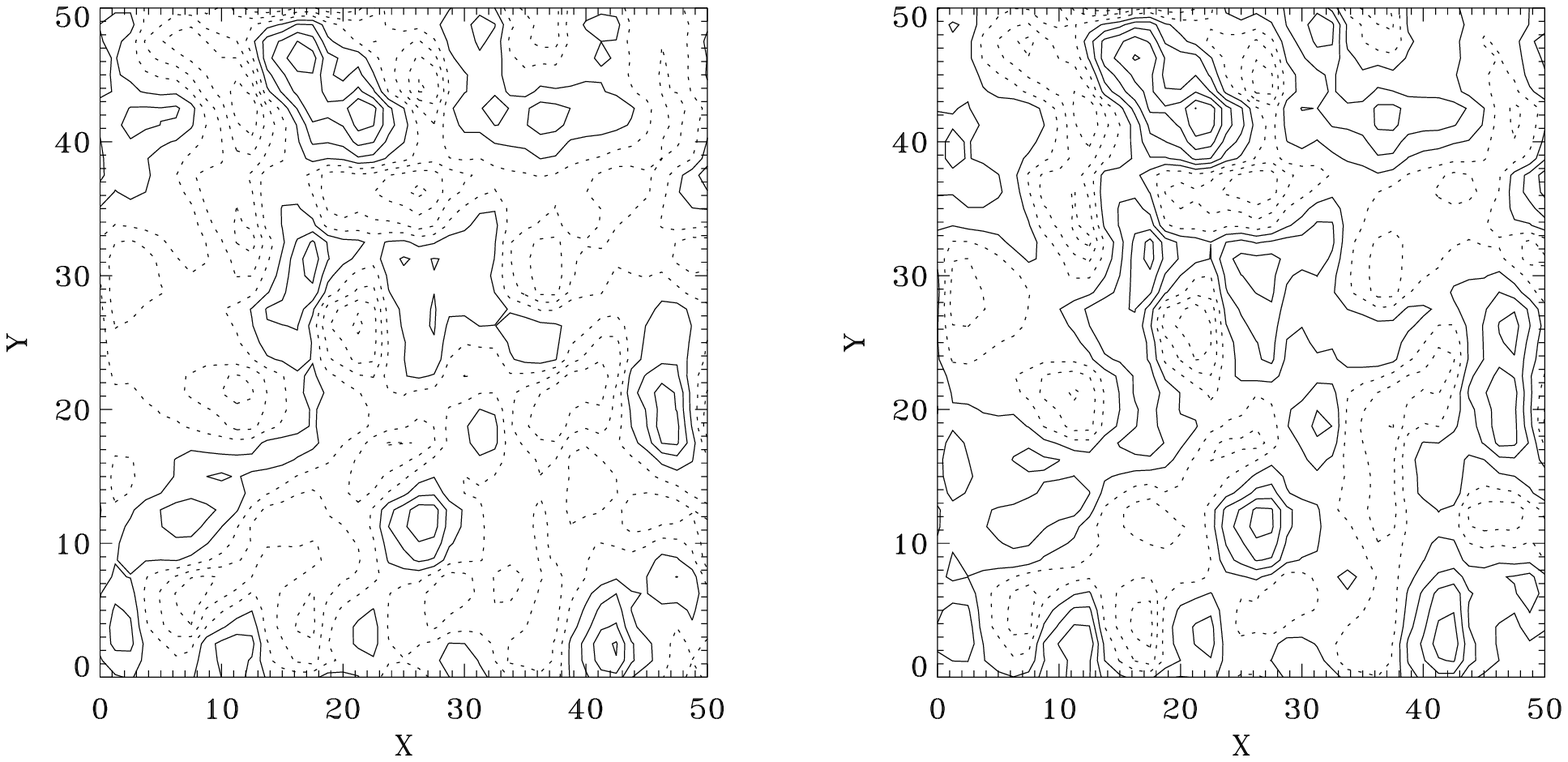,width=9.5cm,height=4.4cm}}
\centerline{\psfig{figure=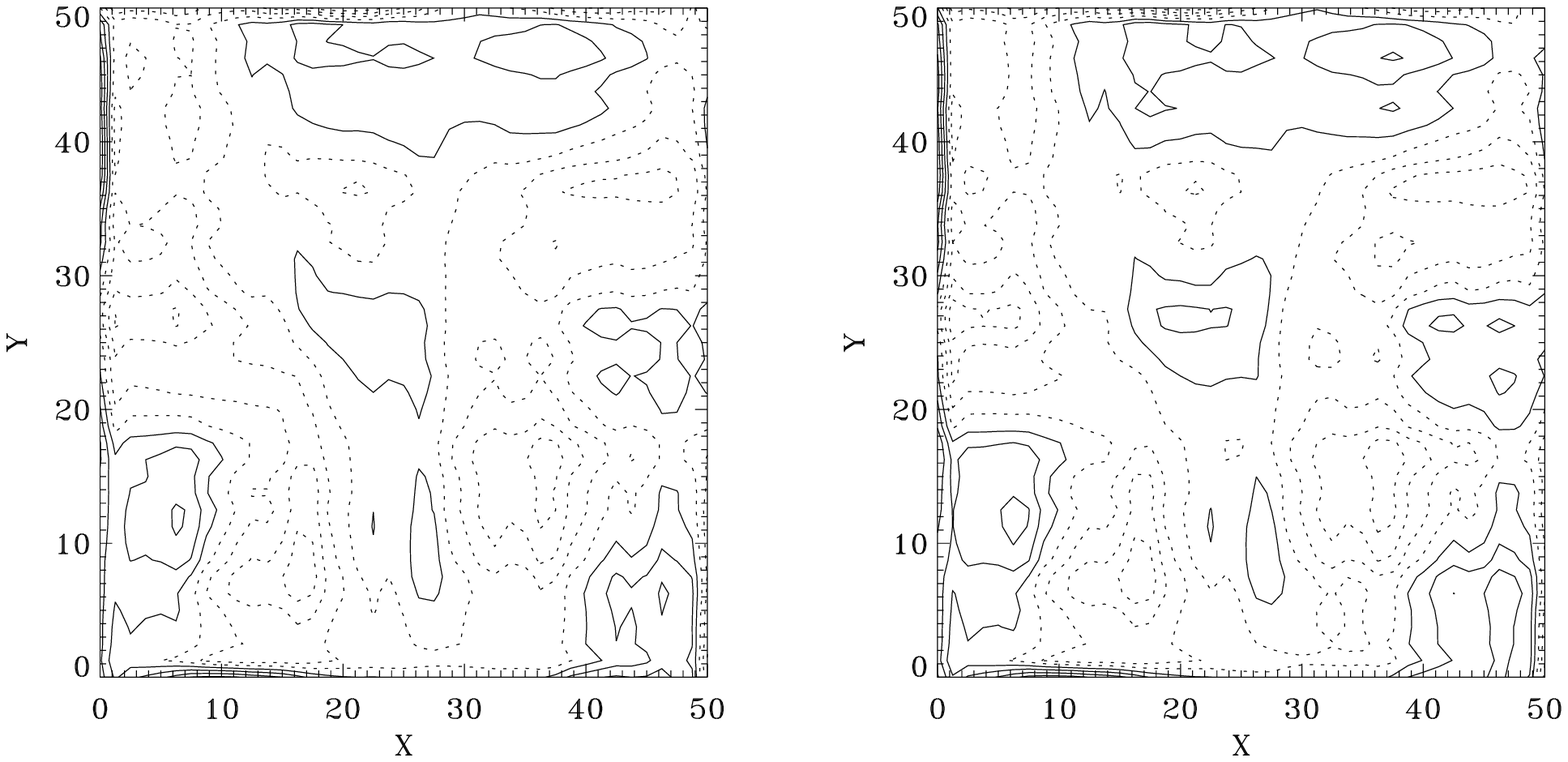,width=9.5cm,height=4.4cm}}
\caption[]{Two views of the divergence field at different resolution. The
difference between the left and right flow fields is the noise introduced
by the Earth's atmosphere. In the first row no filtering has been
applied and common features are barely identifiable. The following
rows show filtered data according to multi-resolution representation
(see Paper I) with a resolution divided by 2 from one row to the next.
The mesh size is 10 pixels (1\farcs25) and the flow is an average over
$\sim1.5$hr. X and Y are in arcsec.}
\label{div_fields}
\end{figure}

\begin{figure}[ht]
\centerline{\psfig{figure=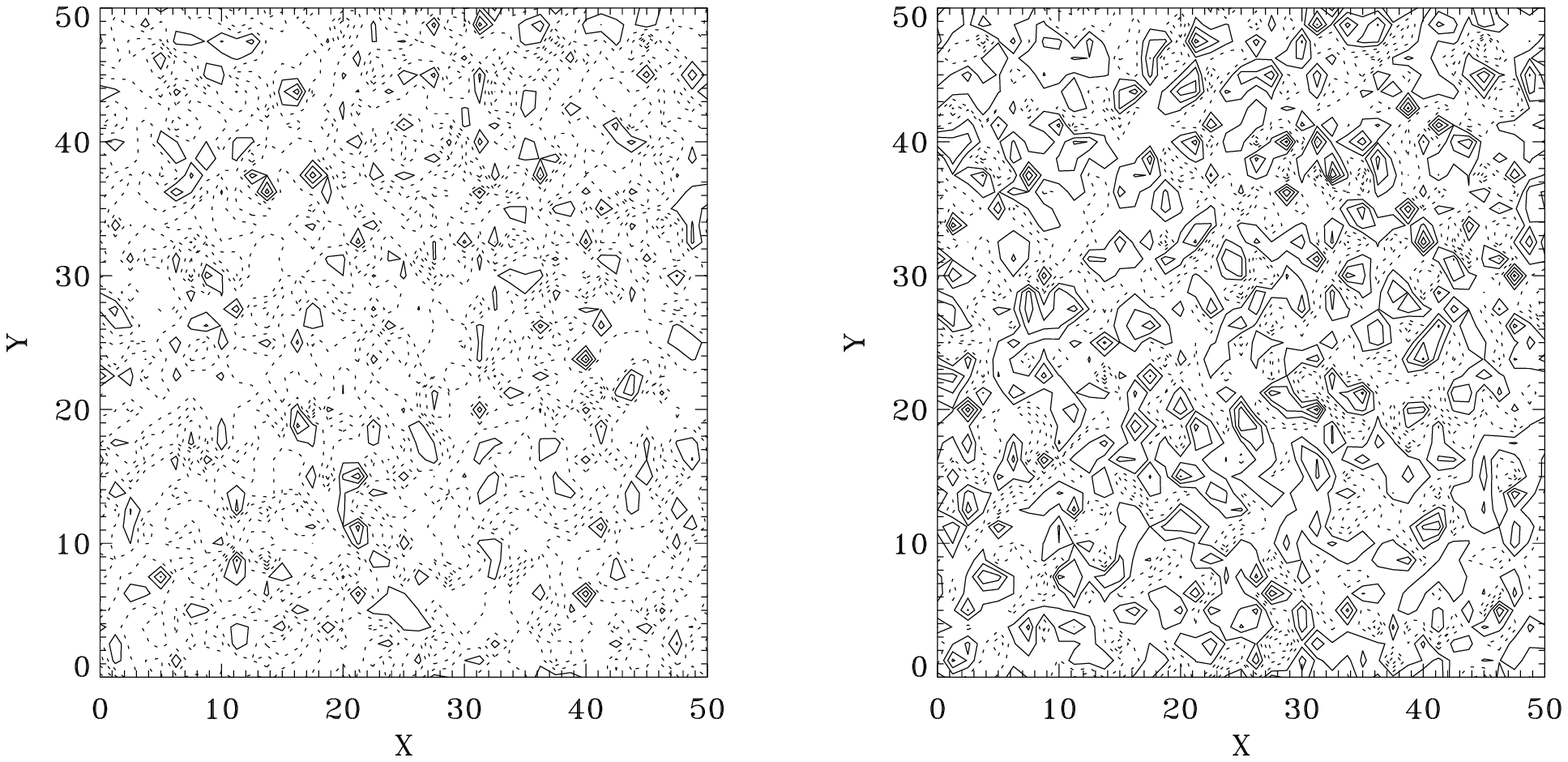,width=9.5cm,height=4.4cm}}
\centerline{\psfig{figure=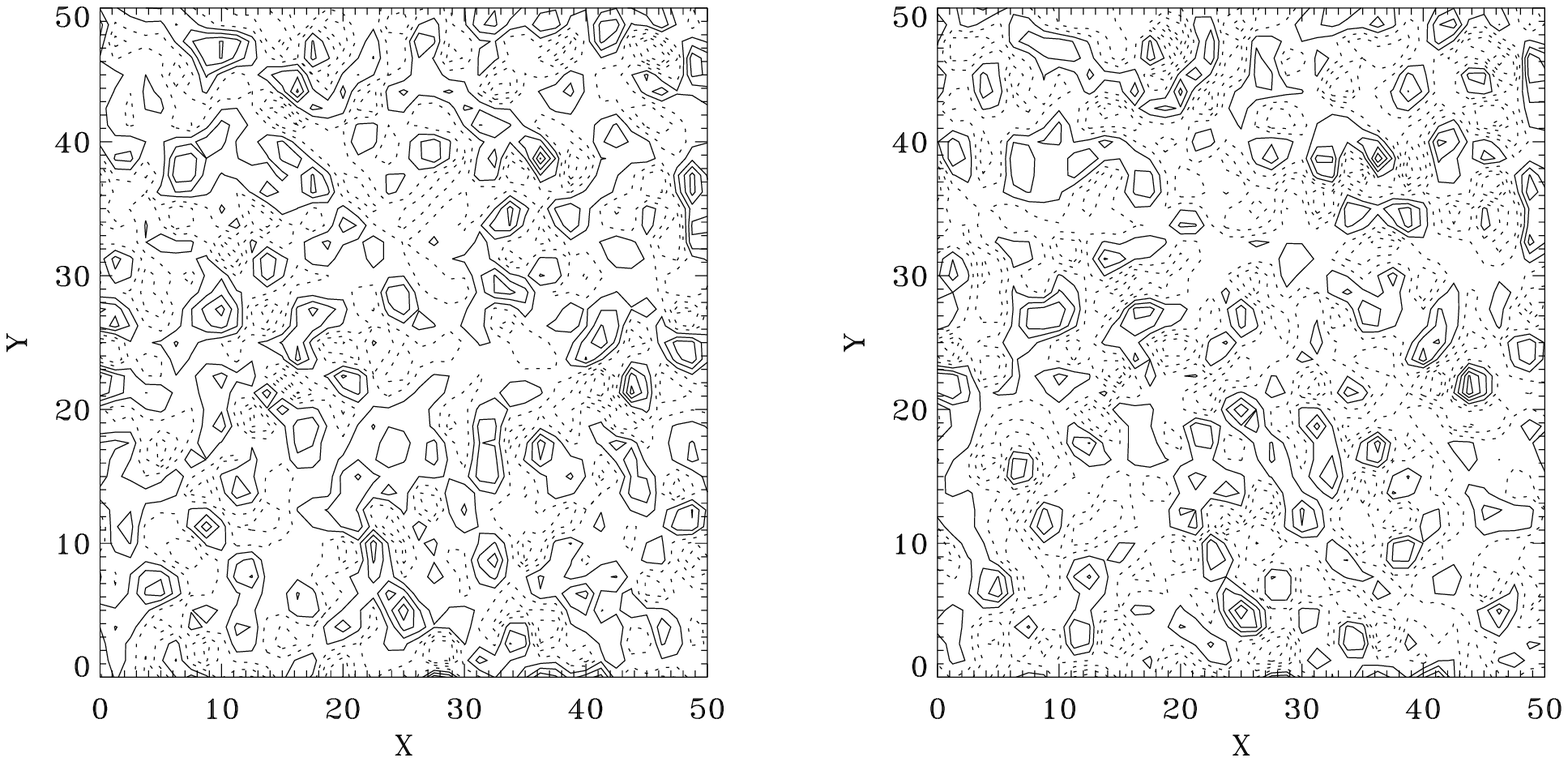,width=9.5cm,height=4.4cm}}
\centerline{\psfig{figure=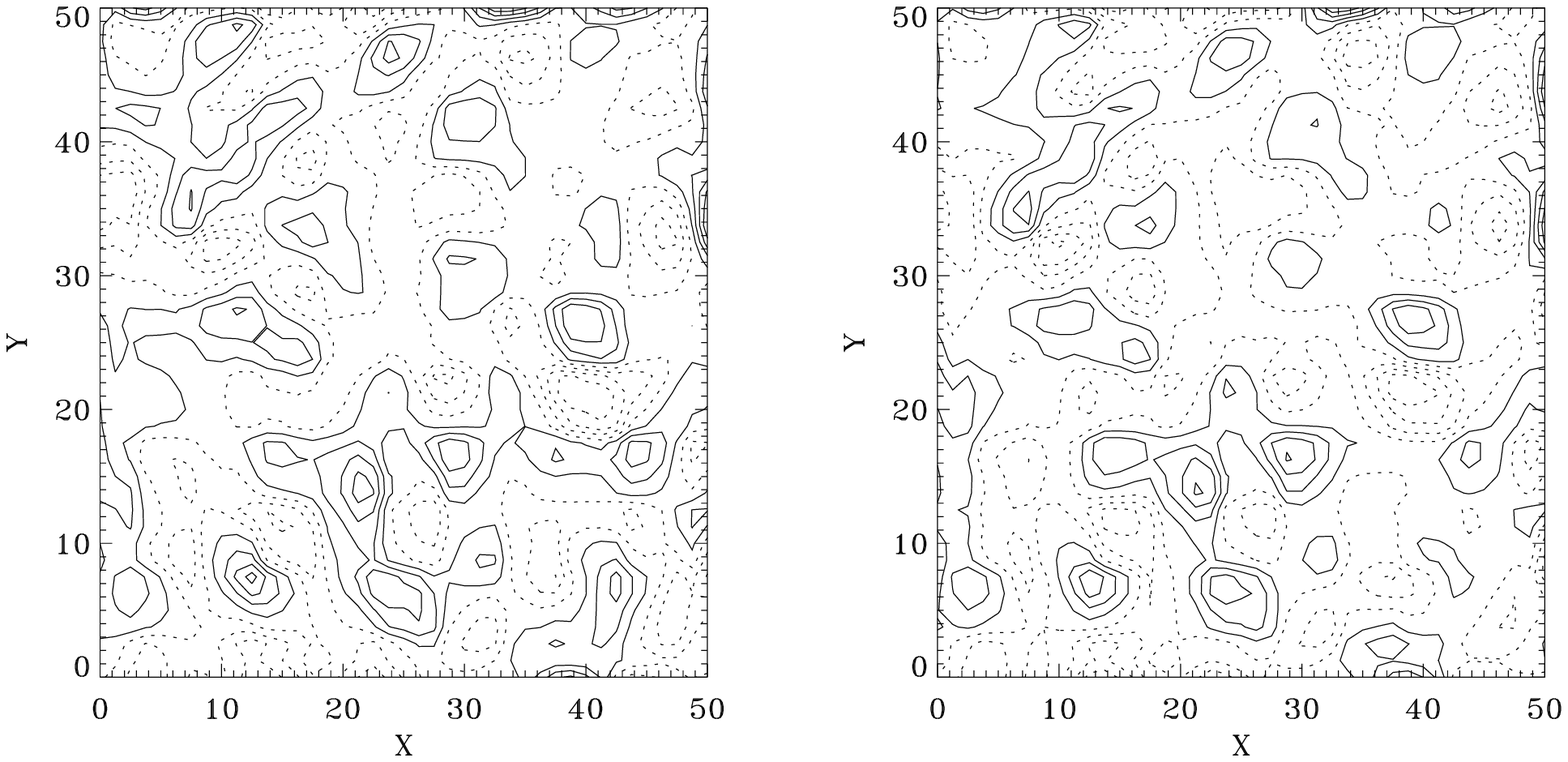,width=9.5cm,height=4.4cm}}
\centerline{\psfig{figure=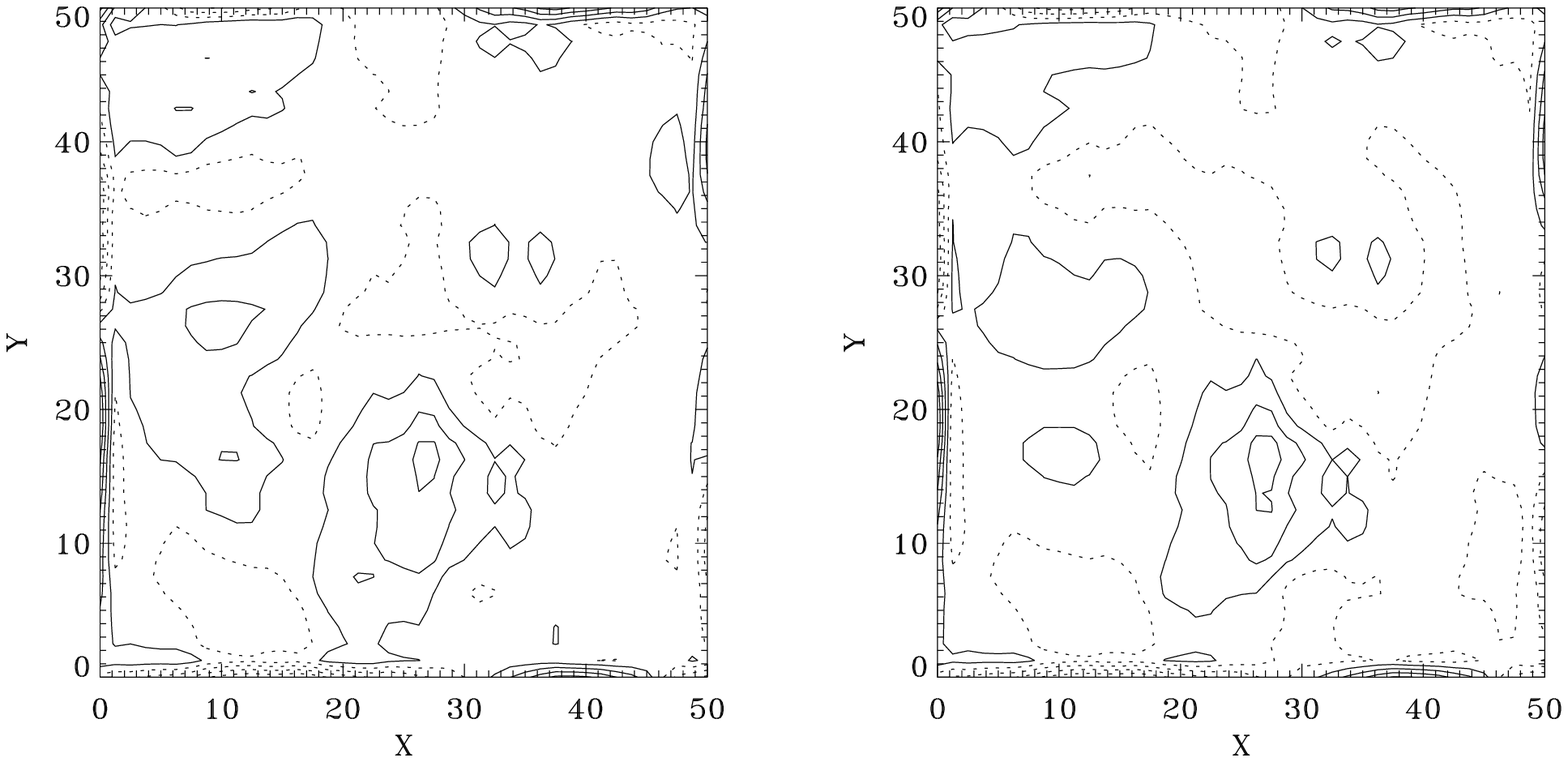,width=9.5cm,height=4.4cm}}
\caption[]{Same as in Fig.~\ref{div_fields} but for the z-component of the
curl field.}
\label{rot_fields}
\end{figure}

\subsection{Influence of the granule lifetime}

However, all granules do not have the same lifetime (see for example
\citealt{HBVH99} or Paper I). In this case, it may be more appropriate to
select only a subsample of the granules. For example, in the simple case
where half of the $N$ granules have a lifetime $\Delta a_t$ and the other
half a lifetime $\Delta b_t=2\Delta a_t$, the error on the velocity is

$$ \sigma_{a+b}^2= \frac{1}{N^2} \left(\sum_{i \in D, \Delta t_i = \Delta a_t} \frac{2\sigma_x^2}{\Delta a_t^2}+ \sum_{i \in D, \Delta t_i = \Delta b_t} \frac{2\sigma_x^2}{\Delta b_t^2}\right) $$
$$ = \frac{1}{N^2} \left(\frac{N\sigma_x^2}{\Delta a_t^2}+
\frac{N\sigma_x^2}{4\Delta a_t^2} \right) =\frac{5\sigma_x^2}{4N\Delta
a_t^2}. $$
If we take only the $\frac{N}{2}$ granules with the largest 
lifetime $\Delta b_t$ into account, we obtain

$$ \sigma_{b}^2=\frac{2^2}{N^2} \sum_{i \in D, \Delta t_i = \Delta b_t} \frac{2\sigma_x^2}{\Delta b_t^2}=\frac{4}{N^2} \frac{N\sigma_x^2}{4\Delta a_t^2}$$
$$=\frac{\sigma_x^2}{N\Delta a_t^2} \le \sigma_{a+b}^2. $$
Therefore, the errors are smaller when considering only the granules with 
the longest lifetime.

This result shows that some selection of ``valid granules" may improve
the quality of the velocity field. We thus impose a lower threshold on
the granule lifetime to eliminate short-lived structures.  This threshold
has to be determined empirically and depends on the filling factor of
the grid. (We need to avoid grid points with no data, see Paper I.)
This strategy may be improved by computing the average velocity for
various thresholds, starting with the highest one. Since granules with
long lifetimes are rare the number $N$ is small and the error is large. By
reducing the threshold, the number of granules increases rapidly and the
error decreases. For some optimal threshold, the error ceases to decrease
as the increase in granule number no longer compensates for the diminishing
value of $\Delta t$ (see Eq.~\ref{preci}).  This process is illustrated
in Fig.~\ref{err_fct_seuil} where we show the dispersion of velocity
differences between the two independent data sets
as a function of the threshold on the granule lifetime. Clearly, for
these data the optimal threshold is around 3~min.

\subsection{Influence of the granule size}

It is also interesting to study the influence of the granule size on the
error for the granule centre of gravity. We would expect a smaller error
in the case of large granules, as these are defined by a larger number
of pixels. A selection on the granule size could then also improve
the precision on the velocity field. Figure~\ref{err_vs_surf} shows an
estimation of the error on the granule position for various granule size
intervals.  The precision increases only slightly with the granule size,
and the improvement is at most 0.05 pixels. This means that the error on the
velocity will not change much either. This small improvement in the error
is too weak to compensate for the increase in the error on the velocity due
to the decrease in the number of granules as their size increases.

\section{Error propagation to the final maps}

\begin{table*}
\begin{center}
\begin{tabular}{ccccccccc}
\hline \hline
\multicolumn{2}{c}{Resolution (pixel size)} & V$_{\rm rms}$ & C$_v$ &
(Div V)$_{\rm rms}$& C$_d$ & (Curl V)$_{\rm rms}$& C$_c$  & Turbulence\\
\hline
1.25\arcsec & 920 km &   700$\pm$695 & 0.49 & 7.1$\pm7.0$ & 0.45 &
5.7$\pm7.0$& 0.22 & 880$\pm$224\\
2.50\arcsec & 1840 &   478$\pm$342 & 0.74 & 3.8$\pm2.8$ & 0.69 &
2.9$\pm3.0$& 0.45 & 893$\pm$122\\
5.00\arcsec & 3680 &   340$\pm$172 & 0.87 & 1.6$\pm0.9$ & 0.89 &
1.2$\pm0.9$& 0.73 & 890$\pm$94\\
10.0\arcsec & 7360 &   205$\pm$92 & 0.90 & 0.58$\pm0.24$ & 0.94
&0.41$\pm0.24$& 0.85 & 896$\pm$65\\
\hline \hline
\end{tabular}
\caption{
Rms velocities, divergences and curls, of the maps issued from the
multi-resolution analysis of the whole time series. The first row is raw
data (no wavelet filtering) while the next rows shows the numbers issued
from the wavelet-filtered maps.  Velocities and turbulence are in m/s,
divergence and curl in $10^{-4}$s$^{-1}$.  }
\label{tabvelo_raw}
\end{center}
\end{table*}

\begin{table*}
\begin{center}
\begin{tabular}{ccccccccc}
\hline \hline
\multicolumn{2}{c}{Resolution (pixel size)} & V$_{\rm rms}$ & C$_v$ &
(Div V)$_{\rm rms}$ & C$_d$
& (Curl V)$_{\rm rms}$ & C$_c$ & Turbulence\\
\hline
1.25\arcsec & 920 km & 597$\pm$330 & 0.85 & 6.8$\pm4.2$& 0.81 &
5.0$\pm4.2$& 0.66 & 680$\pm$193\\
2.50\arcsec & 1840 &   466$\pm$162& 0.94 & 4.0$\pm1.7$ &0.91 &
2.7$\pm1.7$& 0.81 & 692$\pm$105\\
5.00\arcsec & 3680 &   350$\pm$80 & 0.97 & 1.8$\pm0.4$ & 0.97 &
1.2$\pm0.4$& 0.95 & 689$\pm$78\\
10.0\arcsec & 7360 &   209$\pm$40&0.98 &0.63$\pm0.1$ & 0.98
&0.42$\pm0.1$& 0.96 & 694$\pm$53\\
\hline \hline
\end{tabular}
\caption{Same as Table~\ref{tabvelo_raw} but for $k-\omega$ filtered
data. Note the reduction in the dispersion of the results.}
\label{tabvelo}
\end{center}
\end{table*}

\begin{table*}
\begin{center}
\begin{tabular}{ccccccccc}
\hline \hline
\multicolumn{2}{c}{Resolution (pixel size)} & V$_{\rm rms}$ & C$_v$ &
(Div V)$_{\rm rms}$ & C$_d$
& (Curl V)$_{\rm rms}$& C$_c$ &Turbulence\\
\hline
0\farcs875 & 644 km & 722$\pm$534 & 0.73 & 11$\pm$9.7 &0.64 &
10$\pm9.7$& 0.54 & 528$\pm$288\\
1\farcs75  & 1288 &   528$\pm$258 & 0.88 & 6.3$\pm3.7$& 0.83 &
4.5$\pm3.8$& 0.65 & 532$\pm$143\\
3\farcs50  & 2576 &   403$\pm$128 & 0.95 & 2.7$\pm0.9$& 0.94 &
1.8$\pm0.9$& 0.87 & 533$\pm$75\\
7\arcsec   & 5152 &   284$\pm$65  & 0.97 & 1.2$\pm0.3$& 0.97 &
0.7$\pm0.3$& 0.94 & 533$\pm$45\\
\hline \hline
\end{tabular}
\caption{Same as Table~\ref{tabvelo} but for a smaller velocity pixel.}
\label{tabvelo_new}
\end{center}
\end{table*}

Following the algorithm described in Paper I (but see also Sect. 5.1),
we now compute the velocity fields of the two independent time series
at a different resolution (see Paper I). A comparison between the two
fields shows the influence of the distortion induced by the Earth's
atmosphere and the efficiency of MRA in revealing the flow patterns on the
different scales. The results are illustrated by Figs.~\ref{div_fields}
and \ref{rot_fields}, which show the horizontal divergences and the
vertical component of vorticity, respectively. This flow field is the
average velocity over a time lapse of $\Delta t=87$~min including the
preceding sequence and with no distinction between the ``bad'' and
``good'' sequences. Actually, we first computed the velocity fields
using the ``good" sequence and then extended the computation to the
whole set. We observed that the dispersion between the two independent
sequences was still reduced when using the whole set of data showing
that, for the determination of the large-scale mean flow, the observed
degradation in the distortion is not influential.

In Table~\ref{tabvelo_raw}, we quantitatively summarise the dispersion of
the results as a function of the resolution for unfiltered images.
This table has been obtained with the whole time series (87~min),
using a velocity pixel of 1.25\arcsec\ and removing all granules with
a lifetime less than 180~s. We computed the rms velocities, divergences
and curls, of the maps issued from the multi-resolution decomposition.
The ``uncertainties" shown along the numbers give the amplitude of the
fluctuations generated by the Earth's atmospheric distortion.
This table shows the decreasing influence of distortion with increasing
scale. It also shows that, as expected, velocity gradients (divergence
and the z-component of the vorticity) suffer much more from the noise
and that the curl is certainly the quantity most sensitive to image
quality. The values of the correlations between the results issued from
the two independent series (C$_v$, C$_d$ and C$_c$) quantitatively
indicate the similarity of the fields. Here too, filtered fields are
much better correlated, up to 90\% on the velocity field. As shown below,
this correlation is still improved when using $k-\omega$ filtered data.

In this table, we also give a `turbulent velocity', which is the mean
rms dispersion of granule velocities. Indeed, in each velocity pixel
we take the mean velocity of the granules falling in this very pixel;
however, while computing this mean velocity, we also have access to the
dispersion around this mean. This mean (over the whole field of view)
dispersion represents the random motion of granules around their drift by
large-scale flows. This table shows that, although this quantity suffers
(also) from image quality, it is almost independent of scale. This
independence is expected since this quantity measures the proper
motion of granules and therefore their intrinsic kinetic energy, which
should not vary from place to place.

Table~\ref{tabvelo} gives the same quantities but for the $k-\omega$
filtered sequence. We note the strong reduction of the noise, almost a
factor 2, on all scales and the strong improvement of correlation between
the results of the two independent series. Moreover, the dispersion of
the velocities is reduced in the same proportion as displacement
of granules (i.e. $\frac{330{\rm m/s}}{695{\rm m/s}}\sim\frac{0.4{\rm
pix.}}{0.7{\rm pix.}}$). The comparison of the values of turbulence in
Tab.~\ref{tabvelo_raw} and \ref{tabvelo} shows the influence of
the Earth's atmospheric noise on the random motion of granules.

Table~\ref{tabvelo_new} gives another view of this velocity
field using a smaller mesh size. Here we computed the velocity
amplitudes that are traced by granules when they can be considered as
passive scalars, i.e. on a scale larger than 2500~km. Numbers show that
on that scale the velocity field has an amplitude of 400~m/s and that
such a measurement is still uncertain by 30\%. This is about the same
for the divergence, but it rises to 50\% for the vorticity.  The more
intense fluctuations compared to Table~\ref{tabvelo} come from the
smaller scales involved.  The weaker ``Turbulence" values come from the
empty bins which are more numerous.

These results show that, when independent images series are available,
the use of the CST algorithm authorizes a tight control on the role of
the noise induced by the Earth's atmospheric distortion.

\section{Conclusions}

In this paper we have tried to show the various influences of the noise
generated by the Earth's atmosphere when one measures the horizontal
velocity field at the surface of the sun through granule tracking.

For this purpose we compared two time series of images of the solar
surface separated by at most a few seconds; they thus represent the same
solar signal but with a different noise (essentially, distortion from
the Earth's atmosphere). We measured the position of the granules in the
two time series and used these data to determine the amplitude of the
noise. We could thus test the pre-processing that could first be applied
to the images, namely a $k-\omega$ filtering and the destretching. The
comparison between the pre-processed series and the raw one allowed
us to evaluate the efficiency of the pre-processing. It turns out that
$k-\omega$ filtering significantly reduces the noise while destretching,
even if reducing the fluctuations, not only cannot reduce it, but
amplifies it when distortion increases.

We also found that the CST algorithm that tracks coherent structures
(essentially granules) could easily go across sequences of images with
decreased image quality. This is because all granules are not affected
evenly by distortion and granules whose trajectories are too perturbed
by atmospheric noise are eliminated; they thus do not input noise in the
final velocity field. This is clearly a feature that algorithms based on
local correlations of images cannot authorize.

We also studied how the distortion noise affecting granules positions
introduces some noise into the interpolated velocity fields. We thus
showed that the lifetime of granules was an important parameter and
that short-lived granules should be eliminated. The trade-off between
granule number (the more numerous the granules, the better sampled the
velocity field) and noise intensity seems to be, for our sequence, around
3~min. It is clear that, if the atmospheric noise is more intense, this
threshold should be increased. With the decomposition of the velocity
on the Daubechies wavelets (see Paper I), we could evaluate the impact
of the noise on different scales and show that errors on velocities
decrease with increasing scale, as expected. More precisely, we could
show that on a scale of 2500~km, i.e. on the scale where granule motions
trace the large-scale flows \cite[][]{RRLNS01}, the typical velocities,
around 400~m/s, are still noised at a 30\% level. Nevertheless, one can
recognize, on the wavelet-filtered maps, common patterns between the
two time series, all the more easily when the scale is large.

The next steps are now obvious: with new cameras with fast reading sensors
(like CMOS), it is easy to increase the time sampling by a factor 10. In
this case the $k-\omega$ filtering will be much more efficient at reducing
the noise on the granule motion. Hence, with an increased time-sampling we
expect to reduce the noise in two ways: first, by an improved efficiency
of the $k-\omega$ filter and second by a factor $\sqrt{N}$ from the $N$
images sharing the same solar signal.

The tools developed here seem to perform quite efficiently on the
numerical side and therefore allow for the treatment of much larger
fields of view. 

Finally, we did not discuss the results in terms of solar turbulence. Let
us mention that numbers, like the amplitude of velocity on a 2.5~Mm scale,
agree with previous determinations \cite[e.g.][]{BFSSTTT91}. However,
it is clear that (solar) fluid mechanics should be discussed, such as
the scale dependence of flow features. This is beyond the scope of this
paper but will be the subject of forthcoming work.

\begin{acknowledgements}
RT wishes to thank the French ministery of education for its financial
support through the CALAS project.  We are most grateful to Peter Brandt
who kindly allowed us to use the extraordinary eleven-hour sequence
obtained at the SVST of La Palma.
\end{acknowledgements}

\bibliographystyle{aa}

\end{document}